\pdfoutput=1
\documentclass[sigconf]{acmart}

\AtBeginDocument{%
  \providecommand\BibTeX{{%
    \normalfont B\kern-0.5em{\scshape i\kern-0.25em b}\kern-0.8em\TeX}}}

\acmPrice{15.00}
\acmISBN{978-1-4503-XXXX-X/18/06}
\PassOptionsToPackage{hyphens}{url}\usepackage{hyperref}
\usepackage{mdwtab}
\usepackage{multirow}
\usepackage{xspace}
\usepackage{graphicx}
\usepackage{multirow}
\usepackage{tabularx}

\begin{document}

\title{Using AI/ML to Find and Remediate Enterprise Secrets in Code \& Document Sharing Platforms\\
}

\makeatletter
\DeclareRobustCommand\onedot{\futurelet\@let@token\@onedot}
\def\@onedot{\ifx\@let@token.\else.\null\fi\xspace}

\def\eg{\emph{e.g}\onedot} \def\Eg{\emph{E.g}\onedot}
\def\ie{\emph{i.e}\onedot} \def\Ie{\emph{I.e}\onedot}
\def\cf{\emph{c.f}\onedot} \def\Cf{\emph{C.f}\onedot}
\def\etc{\emph{etc}\onedot} \def\vs{\emph{vs}\onedot}
\def\wrt{w.r.t\onedot} \def\dof{d.o.f\onedot}
\def\etal{\emph{et al}\onedot}
\makeatother

\author{Gregor Kerr}
\affiliation{  \institution{JPMorgan Chase}
  \country{} 
}
\email{gregor.kerr@jpmchase.com}

\author{David Algorry}
\affiliation{  \institution{JPMorgan Chase}
  \country{} 
}
\email{david.algorry@jpmchase.com}

\author{Senad Ibraimoski}
\affiliation{  \institution{JPMorgan Chase}
  \country{} 
}
\email{senad.ibraimoski@jpmchase.com}

\author{Peter Maciver}
\affiliation{  \institution{JPMorgan Chase}
  \country{} 
}
\email{peter.maciver@jpmchase.com}

\author{Sean Moran}
\affiliation{  \institution{JPMorgan Chase}
  \country{} 
}
\email{sean.j.moran@jpmchase.com}

\renewcommand{\shortauthors}{Kerr and Algorry, et al.}

\begin{abstract}
We introduce a new challenge to the software development community: 1) leveraging AI to accurately detect and flag up secrets in code and on popular document sharing platforms that frequently used by developers, such as Confluence and 2) automatically remediating the detections (\eg by suggesting password vault functionality). This is a challenging, and mostly unaddressed task. Existing methods leverage heuristics and regular expressions, that can be very noisy, and therefore increase toil on developers. The next step - modifying code itself - to automatically remediate a detection, is a complex task. We introduce two baseline AI models that have good detection performance and propose an automatic mechanism for remediating secrets found in code, opening up the study of this task to the wider community.
\end{abstract}





\received{5 October 2023}
\received[revised]{12 March 2009}
\received[accepted]{5 June 2009}

\maketitle

\section{INTRODUCTION}

Preventing accidental leakage of password credentials, API tokens and private keys within code and on document sharing platforms, such as Confluence, is a critical effort in many organisations. A previous study \cite{meli2019bad} discovered more than 100,000 GitHub repositories with secrets present within source code. Similar types of leaks have resulted in several large scale breaches \cite{awskeys,uberleak,starbucksleak}. Existing tools developed to prevent such breaches leverage heuristic approaches such as regex-based approaches\cite{trufflesecurity,gitleaks,detectsecretsyelp}.
Although these tools provide a great basis for detecting secrets in many environments, due to the nature of these heuristic approaches, substantial noise is generated, leaving users to deal with a large proportion of false positive detections. These false positives are due to the nature of these heuristic approaches as one study \cite{regularexpfalsepositive} suggests the false positives are due to the excessive and faulty regular expressions. There has been some exploration into machine learning approaches \cite{ReduceFalsePositives} to reduce these false positives created from these noisy regular expressions.

We introduce a new challenge to the software development community: firstly, the development of an effective AI/ML model that can accurately identify secrets in code. Secondly, auto-remediate those detections with code that is integrated directly in to replace the detected secret. Finally, we suggest an extension to the detection to free text within collaborative document sharing platforms (DSPs). We describe some baseline AI/ML models, and how a human-in-the-loop learning setup was critical to achieving suitable performance. We suggest two different baseline models for code and natural language. For code we develop a language agnostic machine learning model to predict potential secrets using annotated labels from SMEs. For a DSP we develop an AI/ML model that uses outputs of a heuristic tool to provide weak labels which are then re-annotated to internal gold labels based on the input of SMEs. For auto-remediation of secrets detected in code, we describe an effective openrewrite rules-based solution\footnote{\url{https://docs.openrewrite.org/}}. 
%The properties of different output generation strategies are compared in terms for latency and accuracy.
% \begin{figure*}[!t]
%     \centering
%     \includegraphics[width=\textwidth, scale=0.8]{Cloud Migration CoPilot Architecture_v1.png}
%     \caption{Cloud Migration CoPilot Development Architecture }
%     \label{fig:cmc_architecture}
% \end{figure*}

\section{Approach and Methodology}
For both secrets in code and DSP we leverage the \verb|detect-secrets| library \cite{detectsecretsyelp}. Yelp \verb|detect-secrets|
 is an opensource library for detecting secrets within a codebase~\footnote{\url{https://github.com/Yelp/detect-secrets}}. More specifically, it uses a wide range of plugins each with their own custom heuristics to detect a potential secret. These plugins range from detecting AWS keys, private keys, artifactory credentials to keyword secrets. The result of performing a detect-secrets scan is a baseline file containing secrets found within a given codebase. Due to the nature of some heuristics, the detections can be noisy, introducing a lot of false positive detections that the developer needs to remediate. For this remediation process we leverage \verb|OpenRewrite| which enables developers to successfully automate code maintenance tasks through refactoring of software. Developers are able to remediate potential threats, such as secrets, through accepting generated software that remediates these threats, and consequently merge into their application. It is important that the accepted software is in alignment with the approved set of secure communication patterns within its configuration - resulting in compliant code which can be trusted by developers.

\subsection{Secrets in Code}
 \noindent\textbf{Approach for Code:} 
 The first baseline we construct leverages detect-secrets for finding potential secrets within a codebase. However, due to the nature of this heuristic approach, we find a large amount of false positives within the existing approach - in particular arising from the keyword plugin. The keyword plugin uses a wide range of complex regex's to find patterns (\eg password = somestring or password : somestring) with a vast range of operators to separate the left hand-side variable name to the right hand-side value assignment. However, due to the wide range of regex's used, this is the most noisy plugin when it comes to code. We therefore focus our attention for secrets in code towards the most noisy plugin in yelp \verb|detect-secrets| - keyword. To evaluate the model, we curate a dataset of 10,206 lines of code that \verb|detect-secrets| had labelled as secrets. We then re-annotate these lines to use as gold labels for model evaluation. \noindent\textbf{Featurization for Code:} 
For input to the model we consider 3 main features : the potential secret itself, the extension (\emph{py, java, ini}~\etc) of the file the secret originated from and we calculate the strength of the potential secret by using the entropy of the string.
We leverage \verb|CountVectorizer|~\cite{pedregosa2011scikit} to convert the collection of potential secrets to a matrix of token counts. We use a Logistic Regression \cite{pedregosa2011scikit} model for detection.
Our choice to use a logistic regression model was due to the speed in training and ease of deployment. This model also lends itself to easy explainability of a detection. We employ a 70/10/20 split for training,validation and test set on the 10,206 datapoints (7,622 non-secrets, 2,584 secrets). We optimize for recall when tuning the classifier threshold.
 
\noindent\textbf{Performance for Code:} To test the effectiveness of our model we perform validation on a sample test set size of 1,020 data points containing 261 secrets and 759 non-secrets.

\begin{table}[H]
\tiny{
\centering
\resizebox{\columnwidth}{!}{%
\begin{tabular}{cccc}
\hline
\textbf{\tiny{Technique}}  & \textbf{\tiny{Precision}} & \textbf{\tiny{Recall}}  & \textbf{\tiny{F1}} \\ \hline
\tiny{Yelp Detect-secrets}                     & \tiny{0.26} & \tiny{1} & \tiny{0.41} \\
\tiny{Secret Detection Model}                 & \tiny{0.41} & \tiny{0.99} & \tiny{0.58} 
\end{tabular}%
}}
\caption{Secrets Detection in Code}
\label{tab:coderesults}
\end{table}
In Table \ref{tab:coderesults} we minimally reduce recall by missing one secret, and we gain significantly in other metrics. By reducing false positives, we alleviate the need for developers to remediate 383 false positives out of a possible 759 false detections that were reported by \verb|detect-secrets|. \noindent\textbf{Remediation for Code:}
By using a recipe generation/templating tool such as \verb|OpenRewrite| we can remediate these secrets by consistently generating compliant software for the developer to accept. Our current approach requires a catalog of recipes (a collection of operations to accomplish some higher-level task \cite{OpenRewrite}) for very specific internal use cases for the resolution of these secrets. Although the overall goal of remediating secrets across the software estate through automation is achievable, it requires a significant amount of effort to create a comprehensive set of recipes, given the pattern matching nature of the mechanism. To do this at scale we need a more efficient approach for the identification of patterns and generation of associated recipes. In particular, we look to leverage GenerativeAI to create these recipes for us. %This is particularly true due to our security technologies being predominately proprietary so publicly available recipes are not suitable. %In particular, we look to leverage GenerativeAI to create these recipes for us. %As we look beyond this, it is expected that the use of prompting generative models will also generate the final remediated piece of software. In doing so, will compliment (and may even avoid) the generation and execution of templating and recipes.

\subsection{Secrets in Confluence}
\noindent\textbf{Approach for Document Sharing Platforms (DSPs):} 
For our selected DSP (Confluence), there was no existing solution in place for identifying secrets. Therefore, there was no ground truth on whether a given detection on DSP is a secret or not. Given the scale of the number of pages in the DSP (millions), we opted for a weak labelling approach to train the model. The output of Yelp \verb|detect-secrets| when applied to a subset of the pages was corrected by introducing a re-labelling effort from SMEs in the loop. Our initial dataset spanned over 700,000 pages. \noindent\textbf{Featurisation for DSP:} We masked technical words that refer to the same concept, removed special characters that do not contribute to meaningful information, applied lemmatization and stop word removal. The proportion of the original training data was 1 secret per 100,000 rows of data. To address this imbalance, we created synthetic data based on the most typical values that appear in the secrets, as well as a string created with the exrex library. Using this technique, we added 1,300 new secrets to the training data. We employ Term Frequency-Inverse Document Frequency (TF-IDF) to vectorize the text and transform it into numerical vectors. We chose XGBoost for our machine learning model due to its simplicity, efficiency and ability to handle imbalanced data. 
\noindent\textbf{Performance for DSP:} To test the effectiveness of the XGBoost model, we selected 50,000 web pages. From these pages, we obtained near two million rows, with 67 secrets and nearly 2 million non-secrets, giving a highly imbalanced learning task.  
%\SM{we need to define optimized threshold, precision and recall}
\begin{table}[H]
\tiny{
\centering
\resizebox{\columnwidth}{!}{%
\begin{tabular}{cccc}
\hline
\textbf{Technique} & \textbf{Recall}       & \textbf{Precision}   & \textbf{F1}  \\ \hline
Yelp Detect-secrets                     & 0.15 & 0.45 & 0.22  \\
Secret Detection Model                  & 0.97 & 0.36 & 0.53 
\end{tabular}%
}}
\caption{Secrets Detection in Confluence}
\label{tab:cconfluenceresults}
\end{table}

 In Table \ref{tab:cconfluenceresults}, the precision score is lower with XGBoost, but the recall score is significantly higher. This leads to a considerably lower number of false negatives. By reducing the number of false negatives to only 2 rows, we enable the developers to have greater confidence in detecting almost all secrets in the Confluence pages, without significantly increasing the number of false positives. 

%\begin{figure}[H]
%    \includegraphics[width=\linewidth]{figures/Confmatrix.png}
%    \caption{DSP Model Performance}
%    \label{fig:confmatrixConfluence}
%\end{figure}

\section{Conclusion}
We introduce the task of AI \emph{detection and auto-remediation of secrets} in code and on collaborative document sharing platforms. We present two baseline AI detection models, describing their implementation. We also demonstrate how \emph{OpenRewrite} can be leveraged to make code compliant automatically. We believe there is a significant opportunity for positively impacting this task with large language models (LLMs), particularly in making more larger-scale and diverse changes to code to enforce compliance. 
\section{Disclaimer}
The described work is a prototype and is not a production deployed system.
\clearpage
\bibliographystyle{ACM-Reference-Format}
\bibliography{sample}


\begin{thebibliography}{11}

%%% the Web address.

\ifx \showCODEN    \undefined \def \showCODEN     #1{\unskip}     \fi
\ifx \showDOI      \undefined \def \showDOI       #1{#1}\fi
\ifx \showISBNx    \undefined \def \showISBNx     #1{\unskip}     \fi
\ifx \showISBNxiii \undefined \def \showISBNxiii  #1{\unskip}     \fi
\ifx \showISSN     \undefined \def \showISSN      #1{\unskip}     \fi
\ifx \showLCCN     \undefined \def \showLCCN      #1{\unskip}     \fi
\ifx \shownote     \undefined \def \shownote      #1{#1}          \fi
\ifx \showarticletitle \undefined \def \showarticletitle #1{#1}   \fi
\ifx \showURL      \undefined \def \showURL       {\relax}        \fi
% invisible to TeX
\providecommand\bibfield[2]{#2}
\providecommand\bibinfo[2]{#2}
\providecommand\natexlab[1]{#1}
\providecommand\showeprint[2][]{arXiv:#2}

\bibitem[Ope({[n.\,d.]})]%
        {OpenRewrite}
 \bibinfo{year}{[n.\,d.]}\natexlab{}.
\newblock
\newblock
\urldef\tempurl%
\url{https://docs.openrewrite.org/concepts-explanations/recipes}
\showURL{%
\tempurl}


\bibitem[Basak et~al\mbox{.}(2023)]%
        {regularexpfalsepositive}
\bibfield{author}{\bibinfo{person}{Setu~Kumar Basak}, \bibinfo{person}{Jamison
  Cox}, \bibinfo{person}{Bradley Reaves}, {and} \bibinfo{person}{Laurie
  Williams}.} \bibinfo{year}{2023}\natexlab{}.
\newblock \bibinfo{title}{A Comparative Study of Software Secrets Reporting by
  Secret Detection Tools}.
\newblock
\newblock
\showeprint[arxiv]{2307.00714}~[cs.CR]


\bibitem[Daws(2020)]%
        {starbucksleak}
\bibfield{author}{\bibinfo{person}{Ryan Daws}.}
  \bibinfo{year}{2020}\natexlab{}.
\newblock \bibinfo{title}{Starbucks’ API key found in public github
  repository – reports}.
\newblock
\newblock
\urldef\tempurl%
\url{https://www.developer-tech.com/news/2020/jan/07/starbucks-api-key-found-public-github-repository-reports/}
\showURL{%
\tempurl}


\bibitem[gitleaks(2023)]%
        {gitleaks}
\bibfield{author}{\bibinfo{person}{gitleaks}.} \bibinfo{year}{2023}\natexlab{}.
\newblock \bibinfo{title}{gitleaks}.
\newblock
\newblock
\urldef\tempurl%
\url{https://github.com/gitleaks/gitleaks}
\showURL{%
\tempurl}


\bibitem[Loo(2018)]%
        {detectsecretsyelp}
\bibfield{author}{\bibinfo{person}{Aaron Loo}.}
  \bibinfo{year}{2018}\natexlab{}.
\newblock \bibinfo{title}{Yelp's Secret Detector: Preventing Secrets in Source
  Code}.
\newblock
\newblock
\urldef\tempurl%
\url{https://engineeringblog.yelp.com/2018/06/yelps-secret-detector.html}
\showURL{%
\tempurl}


\bibitem[Meli et~al\mbox{.}(2019)]%
        {meli2019bad}
\bibfield{author}{\bibinfo{person}{Michael Meli}, \bibinfo{person}{Matthew~R
  McNiece}, {and} \bibinfo{person}{Bradley Reaves}.}
  \bibinfo{year}{2019}\natexlab{}.
\newblock \showarticletitle{How bad can it git? characterizing secret leakage
  in public github repositories.}. In \bibinfo{booktitle}{\emph{NDSS}}.
\newblock


\bibitem[Pedregosa et~al\mbox{.}(2011)]%
        {pedregosa2011scikit}
\bibfield{author}{\bibinfo{person}{Fabian Pedregosa}, \bibinfo{person}{Ga{\"e}l
  Varoquaux}, \bibinfo{person}{Alexandre Gramfort}, \bibinfo{person}{Vincent
  Michel}, \bibinfo{person}{Bertrand Thirion}, \bibinfo{person}{Olivier
  Grisel}, \bibinfo{person}{Mathieu Blondel}, \bibinfo{person}{Peter
  Prettenhofer}, \bibinfo{person}{Ron Weiss}, \bibinfo{person}{Vincent
  Dubourg}, {et~al\mbox{.}}} \bibinfo{year}{2011}\natexlab{}.
\newblock \showarticletitle{Scikit-learn: Machine learning in Python}.
\newblock \bibinfo{journal}{\emph{the Journal of machine Learning research}}
  \bibinfo{volume}{12} (\bibinfo{year}{2011}), \bibinfo{pages}{2825--2830}.
\newblock


\bibitem[Powell(2023)]%
        {uberleak}
\bibfield{author}{\bibinfo{person}{Olivia Powell}.}
  \bibinfo{year}{2023}\natexlab{}.
\newblock \bibinfo{title}{Over 77,000 uber employee details leaked online}.
\newblock
\newblock
\urldef\tempurl%
\url{https://www.cshub.com/attacks/news/iotw-over-77000-uber-employee-details-leaked-in-data-breach}
\showURL{%
\tempurl}


\bibitem[Saha et~al\mbox{.}(2020)]%
        {ReduceFalsePositives}
\bibfield{author}{\bibinfo{person}{Aakanksha Saha}, \bibinfo{person}{Tamara
  Denning}, \bibinfo{person}{Vivek Srikumar}, {and}
  \bibinfo{person}{Sneha~Kumar Kasera}.} \bibinfo{year}{2020}\natexlab{}.
\newblock \showarticletitle{Secrets in Source Code: Reducing False Positives
  using Machine Learning}. In \bibinfo{booktitle}{\emph{2020 International
  Conference on COMmunication Systems \& NETworkS (COMSNETS)}}.
  \bibinfo{pages}{168--175}.
\newblock
\urldef\tempurl%
\url{https://doi.org/10.1109/COMSNETS48256.2020.9027350}
\showDOI{\tempurl}


\bibitem[trufflehog(2023)]%
        {trufflesecurity}
\bibfield{author}{\bibinfo{person}{trufflehog}.}
  \bibinfo{year}{2023}\natexlab{}.
\newblock \bibinfo{title}{trufflehog - Find leaked credentials}.
\newblock
\newblock
\urldef\tempurl%
\url{https://github.com/trufflesecurity/trufflehogy}
\showURL{%
\tempurl}


\bibitem[Zorz(2014)]%
        {awskeys}
\bibfield{author}{\bibinfo{person}{Zeljka Zorz}.}
  \bibinfo{year}{2014}\natexlab{}.
\newblock \bibinfo{title}{10,000 github users inadvertently reveal their AWS
  Secret Access Keys}.
\newblock
\newblock
\urldef\tempurl%
\url{https://www.helpnetsecurity.com/2014/03/24/10000-github-users-inadvertently-reveal-their-aws-secret-access-keys/}
\showURL{%
\tempurl}


\end{thebibliography}
%\appendix
\end{document}